\documentclass[USenglish,oneside,twocolumn]{article}

\usepackage[big]{dgruyter_NEW}
\usepackage{amsmath}
\usepackage{amssymb}
\usepackage{amsthm}
\usepackage{stmaryrd}
\usepackage{framed}
\usepackage{color}
\usepackage[ruled,algosection,noend,linesnumbered]{algorithm2e}
\usepackage{subcaption}

\newtheorem{privdef}{Privacy Definition}
\newtheorem{secthm}{Security Theorem}
\newtheorem{insecthm}{Vulnerability Theorem} 
\newtheorem{seclem}{Security Lemma}
\newtheorem*{complem}{Composition Lemma}

\DeclareMathOperator{\arctanh}{arctanh}
 
\begin{document}

 \author[1]{Raphael R. Toledo}

 \author[2]{George Danezis}

 \author[3]{Ian Goldberg}

 \affil[1]{University College London, E-mail: r.toledo@cs.ucl.ac.uk}

 \affil[2]{University College London, E-mail: g.danezis@cs.ucl.ac.uk}

 \affil[3]{University of Waterloo, E-mail: iang@cs.uwaterloo.ca}

 \title{Lower-Cost $\epsilon$-Private Information Retrieval}

 \runningtitle{Lower Cost $\epsilon$-Private PIR}

  \begin{abstract}
{Private Information Retrieval (PIR), despite being well studied, is computationally costly and arduous to scale. We explore lower-cost relaxations of information-theoretic PIR, based on dummy queries, sparse vectors, and compositions with an anonymity system. We prove the security of each scheme using a flexible differentially private definition for private queries that can capture notions of imperfect privacy. We show that basic schemes are weak, but some of them can be made arbitrarily safe by composing them with large anonymity systems.
}
\end{abstract}



\maketitle
\section{Introduction}
Despite many years of research and significant advances, Private Information Retrieval (PIR) still suffers from scalability issues that were identified some time ago~\cite{RaduSion}: both information theoretic~\cite{Chor} (IT-PIR) and computational~\cite{Rep} PIR schemes require database servers to operate on all records for each private query to conceal the sought record. Thus, as the database grows, the time to process each query grows linearly in theory, and super-linearly in practice: as the data processed exceeds the CPU cache, it has to be fetched from the main memory and eventually persistent storage. Furthermore, in IT-PIR each query is processed by multiple PIR servers. As the number of servers increases, so do the communication and computation costs.

Yet the need to privately access large public databases is pressing: for example Certificate Transparency~\cite{CertTrans}, which strengthens TLS certificate issuing, requires clients to look up certificates for sites they have visited, resulting in privacy concerns. The current size of the certificate database precludes trivial downloading of the entire set and requires PIR~\cite{IanCert}, but it cannot scale to the ubiquitous use of TLS in the future. More scalable systems are therefore needed, even at the cost of lowering the quality of protection\footnote{The privacy offered today by Certificate Transparency is simply to have clients download a range of certificates instead of just one. See below for our analysis of this naive dummy requests mechanism.}. Similarly, as the Tor network~\cite{DBLP:conf/uss/DingledineMS04} grows it becomes untenable to broadcast information about all servers to all clients, and a private querying mechanism~\cite{pirtor} will have to be implemented to prevent attacks based on partial knowledge of the network~\cite{DBLP:conf/pet/DanezisS08}.

To address this challenge, we present designs that lower the traditional PIR costs, but leak more information to adversaries. Quantifying that leakage is therefore necessary and we propose a game-based differential privacy definition~\cite{DiffPrivacy} to evaluate and minimize the risk introduced. This definition can also be used to demonstrate the inadequacy of folk designs for private queries: in the first design, a client queries an untrusted database by looking up the sought record along with other `dummy' records~\cite{balsa2012ob,hong2004architecture,kido2005anonymous} to confuse the adversary; in the second design, a client fetches a record from an untrusted database through an anonymity system~\cite{DBLP:conf/uss/DingledineMS04,ghinita2007prive} to hide the correspondence between the client and the server.

\vspace{3mm}
\noindent The contributions of this paper are:
\begin{itemize}
\item We present and motivate a flexible differential privacy-based PIR definition, through a simple adversary distinguishability game, to analyze lighter-weight protocols and estimate their risk. This is necessary to quantify leakage, but can also capture systems that are arbitrarily private, including computationally and unconditionally private schemes.
\item We argue that simple private retrieval systems using dummies and anonymous channels are not secure under this definition. A number of proposed systems have made use of such private query mechanisms, and we show they can lead to catastrophic loss of privacy when the adversary has side information.
\item We present a number of variants of PIR schemes that satisfy our definition, and compare their security and cost. Our key result is that their composition with an anonymity system can achieve arbitrarily good levels of privacy, leading to highly secure alternatives to traditional PIR schemes.
\item As a means to achieving the above we present a generic composition theorem of differentially private systems with anonymity channels, which is of independent interest.
\item We present an optimization to reduce PIR costs and speed up an IT-PIR scheme by contacting fewer databases, and evaluate it using a Chor-like scheme as an example.
\end{itemize}

The rest of the paper is organized as follows. We present related work on PIR, anonymity systems, and the uses of differential privacy in the remainder of this section. After introducing the paper's notations, we define the threat model and present the privacy definitions in section~\ref{sec:definitions}. We then demonstrate why the use of dummies and anonymity system alone does not guarantee privacy under our definitions in section~\ref{sec:nonepriv} and present our $\epsilon$-private designs and an optimization to cut the computation cost in sections~\ref{sec:epriv} and \ref{sec:optimizing}. Finally in section~\ref{sec:comparison}, we discuss the costs and efficiency of the designs before concluding the paper in section~\ref{sec:conclusions}.

\subsection{Related Work}

Private Information Retrieval (PIR) was introduced in 1995 by Chor et al.~\cite{Chor} to enable private access to public database records. As initially proposed, PIR provided information-theoretic privacy protection (IT-PIR) but required multiple non-collaborating servers each with a copy of the database.  Later, Computational PIR (CPIR)~\cite{Rep} was proposed using a single server, but its practicality has been questioned as being slower than simply downloading the entire database at typical network speeds~\cite{RaduSion}.
Since that time, however, newer CPIR schemes that are significantly faster than downloading the entire database have been proposed~\cite{AG07,XPIR,OG11}.
While IT-PIR offers perfect privacy---the confidentiality of the query cannot be breached even with unlimited resources---it is still a computational burden on multiple databases, since all records must be processed for each query and by each server. 
IT-PIR has been gradually enhanced over time with new capabilities, such as batch processing~\cite{Batch}, multi-block queries~\cite{IanBlock} and tolerance to misbehaving servers~\cite{BS02}. Alternative approaches to scaling PIR include using trusted hardware elements~\cite{DBLP:conf/pet/AsonovF02}.

Research on Anonymity Systems (AS) began in 1981 by David Chaum introducing the mixnet for anonymous e-mail \cite{Pseudo}. Other AS applications were then studied, such as peer-to-peer and web browsing, in particular in the context of Onion Routing and Tor~\cite{DBLP:conf/uss/DingledineMS04}. The Anonymity System accepts a batch of messages and routes them while hiding the correspondence between senders and receivers. Low-latency anonymity systems, however, may still fail under attacks such as traffic analysis and fingerprinting~\cite{Fingerprinting}. Cascade mix networks offer perfect privacy at the cost of lower performance \cite{berthold2001disadvantages}. In this work we will always consider an ideal anonymity system that can be abstracted, from the perspective of an adversary, as an unknown random permutation of input messages to output messages. Real-world instantiations are imperfect and security parameters may have to be adapted, or in the case of onion routing systems~\cite{DBLP:conf/uss/DingledineMS04} some additional batching and mixing may be required.

Differential Privacy definitions and mechanisms were first presented in 2006~\cite{DiffPrivacy} to enable safe interactions with statistical databases.  However, this definition has since been used in machine learning~\cite{DeepLearning}, cloud computing~\cite{Airavat}, and location indistinguishability together with PIR~\cite{GeoIndis} to evaluate and minimize the privacy risk. Differentially private definitions have several advantages: they capture the intrinsic loss of privacy due to using a mechanism, and they are not affected by side information the adversary may hold. Well-known composition theorems can be applied. We note that Chor et al.~\cite{Chor} also make passing allusion to statistical and leaky definitions of PIR in their seminal paper, only to focus on perfectly information-theoretic schemes.

\section{Definitions and $\epsilon$-Privacy}
\label{sec:definitions}

In this work we characterize as PIR any system where a user inputs a secret index of a record in a public database, and eventually is provided with this record, without a defined adversary learning the index. We note that the systems we propose use different mechanisms from traditional IT-PIR and CPIR, and make different security assumptions. Yet they provide the same functionality and interface, and in many cases can be used as drop-in replacements for traditional PIR.

\subsection{Notation}

\textbf{Entities.}
All systems we explore allow $u$ users $\mathcal{U},\ i \in  \llbracket 1,u \rrbracket,$ to perform $q$ queries by sending $p$ requests to the database system $\mathcal{DS}$. 
A database system $\mathcal{DS}$ is composed of $d \in \mathbb{N}$ replicated databases $\mathcal{DB}_{i \in  \llbracket 1,d \rrbracket}$. Each of them stores the same $n$ records of standardized size $b$ bits.
We assume a cascade mix network provides an anonymous channel all users can access. We abstract this Anonymity System as one secure sub-system providing a perfectly secret bi-directional permutation between input and output messages. 

When presenting mechanisms not using the anonymity system we will simply present the interactions of a single user with the database servers, and assume that all user queries can be answered by trivial parallel composition. However, when reasoning about PIR systems using an anonymity system, all user queries are assumed to transit though the anonymous channel.

\vspace{3mm}
\noindent\textbf{Costs.}
This work studies PIR scalability, and we focus on analyzing costs on the server side, which is the performance bottleneck of current techniques. We denote the communication cost as $C_{m}$ corresponding to the number of record blocks sent to the user by $\mathcal{DS}$. The computation cost $C_p$  corresponds to the sum, for each record accessed, of the record access cost and the processing cost by the servers (e.g., the number of XORs), $C_p = N_{record \ access}\cdot (c_{acc}+c_{prc})$.

\subsection{Privacy Definition}

\noindent\textbf{Threat Model.} We consider an adversary $\mathcal{A}$ has corrupted $d_{a}$ databases, out of $d$, in order to discover the queries of a target user $\mathcal{U}_{t}$. These corrupted servers passively record and share all requests they observe to achieve this objective. We also assume that the adversary observes all the user's incoming and outgoing communication. However, in all presented systems, the users encrypt their requests with the servers' public keys, and we assume that for communication with honest servers, only message timing, volume and size are ultimately visible to the adversary. Similarly, using standard mix cascade techniques~\cite{berthold2001disadvantages}, we assume the adversary cannot distinguish the correspondences of input and output messages through an anonymity system. We also assume that the other $u-1$ users in the system are honest, in that they will not provide the adversary any of the secrets they generate or use. However, the adversary also knows the distribution of their queries---an assumption that is necessary to model attacks based on side or background information.

\vspace{3mm}
\noindent\textbf{Security Definitions.} We define ($\epsilon$,$\delta$)-privacy as a flexible privacy metric to fully capture weaker as well as strong privacy-friendly search protocols. The definition relies on the following game between the adversary and honest users: an adversary provides a target user $\mathcal{U}_{t}$ with two queries $Q_i$ and $Q_j$, and to all other users, $\mathcal{U}\setminus \mathcal{U}_{t}$, a single distinct query $Q_{0}$. The target $\mathcal{U}_{t}$ selects one of the two queries, and uses the PIR system in order to execute it, and all others execute $Q_0$, leading to the adversary observing a trace of events $\cal{O}$. This trace includes all information from corrupt servers and all metadata of encrypted communications from the network. We then define privacy as follows:

\begin{privdef}
A protocol provides \emph{($\epsilon ,\delta$)-private PIR} if there are non-negative constants $\epsilon$ and $\delta$, such that for any possible adversary-provided queries $Q_i, Q_j$, and $Q_0$, and for all possible adversarial observations ${\cal{O}}$ in the observation space $\Omega$ we have that
$$\displaystyle \Pr( {\cal{O}} | Q_{i}) \leq  e^{\epsilon} \cdot \Pr( {\cal{O}} | Q_{j}) + \delta.$$
\end{privdef}

In the important special case where $\delta = 0$ we call the stronger property $\epsilon$-privacy, and can also define the likelihood ratio ${\cal L}$:

\begin{privdef}
The likelihood ratio of a particular observation $\cal{O}$ in an \emph{$\epsilon$-private PIR} scheme is defined by:
\( \displaystyle \Pr( {\cal{O}} | Q_{i}) / \Pr( {\cal{O}} | Q_{j}) \leq e^{\epsilon} \),
and the likelihood ratio of the scheme itself is the maximal such value:
$$ \displaystyle \mathcal{L} = \max_{Q_i,Q_j,Q_0,\cal{O}} \frac{\Pr( {\cal{O}} | Q_{i})}{\Pr( {\cal{O}} | Q_{j})} \leq e^{\epsilon}.$$
\end{privdef}

These definitions combine aspects of game-based cryptographic definitions and also differential privacy. We first note how the target user $\mathcal{U}_t$ may chose either $Q_i$ or $Q_j$ with arbitrary a-priori probability, rather than at random. The prior preference between those does not affect $\Pr( {\cal{O}} | Q_{i})$ or $\Pr( {\cal{O}} | Q_{j})$ that relate to the quantity to be bounded, making this definition independent of the adversary's prior knowledge of the target user's query. 

Similarly the defined security game assumes that the adversary knows precisely the queries of all users except the target ($\mathcal{U} \setminus \mathcal{U}_t$), thus capturing any susceptibility to side information they would have about the queries of other users. We note that while users are provided with adversarial queries, the adversary does not learn either any user secrets created as part of the PIR protocols or the user requests sent to honest database servers. 

\vspace{3mm}
\noindent\textbf{Generality and necessity of definition.} In the preferable case $\delta = 0$, the likelihood ratio of any observation is bounded, and we can therefore cap privacy leakage in all cases. A non-zero $\delta$ denotes cases where the leakage may be unbounded: events catastrophic to privacy may occur with probability at most $\delta$. In those cases, requiring $\delta$ to be a negligible function yields the traditional computational cryptographic definitions.

In the case $\epsilon=0$, we recover the cryptographic definition of indistinguishability. The traditional unconditional security provided by IT-PIR is equivalent to a mechanism with $\epsilon = 0$. For $\epsilon > 0$ information about the query selected leaks at a non-negligible rate, and users should rate-limit recurring or correlated queries as for other differentially private mechanisms~\cite{DiffPrivacy}.

Thus we lose no generality by using this definition: it can capture information-theoretic PIR systems, computational PIR systems, as well as systems that leak more information. In the rest of the paper we will define such leaky systems, making this relaxed definition necessary; we will also show that the composition of an $\epsilon$-private PIR mechanism with an anonymity system can lead to systems that provide arbitrarily good privacy.

As for the original differential privacy definition, the $\epsilon$-private PIR definition (with $\delta=0$) ensures that there is no observation providing the adversary certainty to include or exclude any queries from the a-priori query set. When a PIR system does not provide such a guarantee there exist observations that allow the adversary to exclude candidate queries with certainty, leading to poor composition under repeated queries, as studied in the next section. Furthermore, the composition of non $\epsilon$-private PIR schemes with an anonymity channel is not guaranteed to approach perfect privacy as it may leak a lot of information to an adversary with side information about the target, or knowledge about the queries performed by other users.

\section{Non $\epsilon$-Private Systems}
\label{sec:nonepriv}

In this section we analyze two approaches to achieving query privacy that we show are not $\epsilon$-private. We also examine their properties for extreme and impractical security parameters as well as when they are composed with an anonymity system. 

We note that the literature does not refer to those as ``Private Information Retrieval'' or PIR, reserving this term for information theoretically and computationally secure schemes. Yet these ad-hoc systems fulfill the same privacy and functional role as PIR: they are used to lookup a record privately out of a public database, at a lower cost compared with IT-PIR or CPIR. Thus we examine them, analyze their properties, and use some of their ideas as ingredients to build more robust low-cost systems.

\subsection{Naive Dummy Requests} A number of works attempt to hide a true user query to a single untrusted database, by hiding it among a number $p$ of artificially generated user queries (`dummies') to achieve some privacy; for example OB-PWS~\cite{balsa2012ob} in the context of web search, and Hong and Landay~\cite{hong2004architecture} and Kido et al.~\cite{kido2005anonymous} in the context of private location queries. Zhao et al.\ propose a dummy-based privacy mechanism for DNS lookups~\cite{DBLP:conf/mue/ZhaoHS07}, but Hermann et al.\ find its security lacking~\cite{DBLP:conf/sec/HerrmannMF14}. It is interesting to note that both location and DNS applications involve large databases making traditional PIR prohibitively expensive. We show that this mechanism is not $\epsilon$-private, leading occasionally to spectacular information leaks as reported.

\begin{algorithm}
\SetKwComment{Comment}{//}{}
\DontPrintSemicolon
\KwIn{Query $Q \quad (0 \leq Q < n)$; \\ \qquad Security parameter $p \quad (p > 1)$;}
$Req \gets \{ Q \}$;\\
\While { $|Req| < p$}
{
	$Q' \gets \textbf{random}(n)$;\\
	$Req \gets Req \cup {Q'}$;
}
\ForAll{$r\in Req$}
{
	$(index_r, rec_r) \gets \textbf{sendreceive}(\mathcal{DS}, r)$;
}
\KwRet $rec_Q$;
\caption{Naive Dummy Requests (User)}
\end{algorithm}
The function $\textbf{random}(n)$ samples uniformly an integer from $0$ to $n-1$, and $\textbf{sendreceive}(D, m)$ sends a message $m$ to $D$, and returns any response from $D$.

\begin{insecthm}
The Naive Dummy Requests mechanism for security parameter $p < n$ is not $\epsilon$-private.
\end{insecthm}
\begin{proof}
The adversary controlling the database observes which records are queried. Without loss of generality, in case the user queries for $Q_i$, with some probability $\mathcal{A}$ does not see the query requesting record $j$ : $Q_{j}$. We denote by $\Pr( {\cal{O}} | Q_{i})$ the probability an adversary $\mathcal{A}$ observes a trace of events $\mathcal{O}$ knowing the query $Q_{i}$ was sent. Thereby, as the adversary has not seen the query $Q_j$ in the current observation $\cal{O}$, the adversary knows the record $j$ was not sought by the user, hence $\Pr( {\cal{O}} | Q_{j})=0$. Consequently, there is no $\epsilon$ such that $\Pr( {\cal{O}} | Q_{i}) / \Pr( {\cal{O}} | Q_{j}) \leq  e^{\epsilon}$. As the $\epsilon$-privacy bound must apply for any observation $\mathcal{O}$,  and requests $Q_i$ and $Q_j$, this counter example shows that the use of dummies alone does not guarantee $\epsilon$-privacy.
\end{proof}

Practically, this means that if $p < n$, the adversary observing the database system $DS$ will be able to learn, with perfect certainty, that records that have not been requested are not the sought record $Q$. Thus, this mechanism is not $\epsilon$-private, until $p = n$ at which point it becomes perfectly private ($\epsilon = 0$) and corresponds to the naive download of the full database. 

This weakness has practical implications: in the case of a location privacy mechanism an adversary learns which locations a user is not in with certainty, and in the context of DNS lookups, which domains are not being requested. If using this naive scheme in the context of DP5~\cite{DBLP:journals/popets/BorisovDG15}, a system using PIR to protect users' social networks, an adversary would learn with certainty which social links are not present at each query.

\subsection{Naive Anonymous Requests} 

Sending a query through an anonymity system has been proposed to maintain privacy against an untrusted database: the seminal Tor system \cite{DBLP:conf/uss/DingledineMS04} supports private queries to websites, but also performs anonymous requests as a way to resolve \emph{dot-onion} addresses to rendezvous points. Priv\'e~\cite{ghinita2007prive} uses an anonymity system to query location-based services, and another proposal to perform private search engine queries~\cite{DBLP:conf/wpes/Saint-JeanJBF07}. However, this technique alone does not provide $\epsilon$-private PIR.

\begin{algorithm}
\SetKwComment{Comment}{//}{}
\DontPrintSemicolon
\KwIn{Query $Q \quad (0 \leq Q < n)$;}
$(index_Q, rec_Q) \gets \textbf{anonsendreceive}(DS, Q)$; \\
\KwRet $rec_Q$;
\caption{Naive Anon.\ Request (User)}
\end{algorithm}

In this mechanism users simply send requests for the records they seek to the database service through a bi-directional anonymity channel, allowing for anonymous replies (the $\textbf{anonsendreceive}$ function). Upon receiving an anonymous request for a record, the database server simply sends the record back to the user though the anonymous channel. The hope is that since multiple queries are mixed together, the exact query of the target user is obscured. However, there is significant leakage and the mechanism in not $\epsilon$-private.

\begin{insecthm}
The Naive Anonymous Requests mechanism is not $\epsilon$-private, for any number of users $u$ using the system.
\end{insecthm}
\begin{proof}
Following our game-based definition for $\epsilon$-privacy non-target clients' queries are provided by the adversary and are all $Q_0$. Thus the adversary will observe one of $Q_i$ or $Q_j$ only, and all other requests will be for $Q_0$. For the record $Q_x, x \in \{i,j\}$ that is not queried, the probability $\Pr( {\cal{O}} | Q_x)$ equals zero, and the likelihood ratio $\mathcal{L}$ goes to infinity. Thus there exists no $\epsilon$ that may bound this likelihood.
\end{proof}

The proof relies heavily on the fact that the adversary provides all non-target users with a known query $Q_0$ and is therefore able to filter those out at the corrupt database server, and uncover trivially the target user's query. This is an extreme model; however, it also covers realistic attacks. For example, if the adversary knows that most other users are not going to access either $Q_i$ or $Q_j$, but suspects that the target user might, a single observation can confirm this suspicion. This could be the case, for example, when users attempt to look up unpopular, or even personal records that only concern, and are accessed by, the target. The fact that the security parameter of the system, namely the number of users, does not affect security is particularly damning.

\subsection{Composing Naive Mechanisms}

Interestingly, the composition of the two naive mechanisms, namely when multiple users perform Naive Dummy Requests through an anonymous channel, for any $p > 1$, the mechanism becomes $(\epsilon, \delta)$-private. 
This simply involves replacing the $\textbf{sendreceive}$ method with an anonymous channel $\textbf{anonsendreceive}$ in the Naive Dummy Requests algorithm. As the number of users $u$ increases the probability $\delta$ any record is requested zero or $u$ times exactly becomes negligible and then there exists an $\epsilon$ that satisfies the definition.

More specifically, in our indistinguishability game scenario, the probability the adversary observes exactly $u$ queries $Q_i$ is bounded above by $\delta_u \le \left ( \frac{p-1}{n-1}\right )^{u-1}$ while the probability they receive no $Q_i$ queries is bounded above by $\delta_0 \le \left ( \frac{n-p}{n-1}\right )^{u-1}$. (The proof can be found in Appendix~\ref{app:naivecompproof}.) This requires a large number of users $u$ or volume of dummies $p$ to provide meaningful privacy against the single corrupt server. For this reason we instead explore multi-server mechanisms in the next sections. 

\section{Four $\epsilon$-Private PIR Systems}
\label{sec:epriv}
\subsection{Direct Requests}
\label{sec:directreq}

The first $\epsilon$-private PIR mechanism uses dummy queries on \emph{multiple} PIR databases, of which $d_a$ are adversarial and $(d - d_a)$ are honest. The user generates a query for the sought record, along with $p-1$ random (distinct) other ones. The requests are partitioned into sets of equal size and sent to the PIR databases directly. Each database then responds with the list of records requested, encrypted as are all communications.

\begin{algorithm}
\SetKwComment{Comment}{//}{}
\DontPrintSemicolon
\KwIn{\\
\qquad $Q$: $\quad (0 \leq Q < n)$; \\ 
\qquad $p$: $\quad (p > 1) \wedge p \equiv 0 \mod d $;}
$Req \gets \{ Q \}$;\\
\While { $|Req| < p$}
{
	$Q' \gets \textbf{random}(n)$;\\
	\If{$Q' \not \in Req$}
	{$Req \gets Req \cup {Q'}$;}
}

\For{$1 \le i \le d$}
{
	\For{$1 \le j \le p / d$}
	{
		$r \gets \textbf{pop}(Req)$
		$(index_r, rec_r) \gets \textbf{sendreceive}(DB_i, r)$;
	}
}
\KwRet $rec_Q$;
\caption{Direct Requests (User)}
\end{algorithm}

The database servers simply respond to requests by returning the index and the records sought over the encrypted channel. $\textbf{pop}(Req)$ returns items from the set $Req$ independently of the order they were inserted; for example, it could return the smallest item (and also removes it from $Req$). $rec_{Q}$ is the sought record of index $Q$.

\begin{secthm} 
The \emph{direct requests mechanism} is an $\epsilon$-private PIR mechanism with
$$\epsilon = \ln \left ( \frac{1}{d-d_a}\cdot \left ( d \cdot \frac{n-1}{p-1} - d_a\right ) \right ),$$
where $d$ is the number of databases, of which $d_a$ are adversarial, $n$ is the total number of records, and $p$ is the total number of queries sent by the user.
\end{secthm}
\begin{proof}
See Appendix~\ref{app:proofthm1}.
\end{proof}

\noindent\textbf{Costs.} 
For the Direct Request algorithm, as $p$ records are requested and sent back to the user the communication cost is $C_m=p $. As the databases do not XOR any records but just accesses and sends them, the computation cost is  $C_{p}=p \cdot c_{acc}$. \\

\begin{figure}[t]
\centering 
\includegraphics[width=0.5\textwidth]{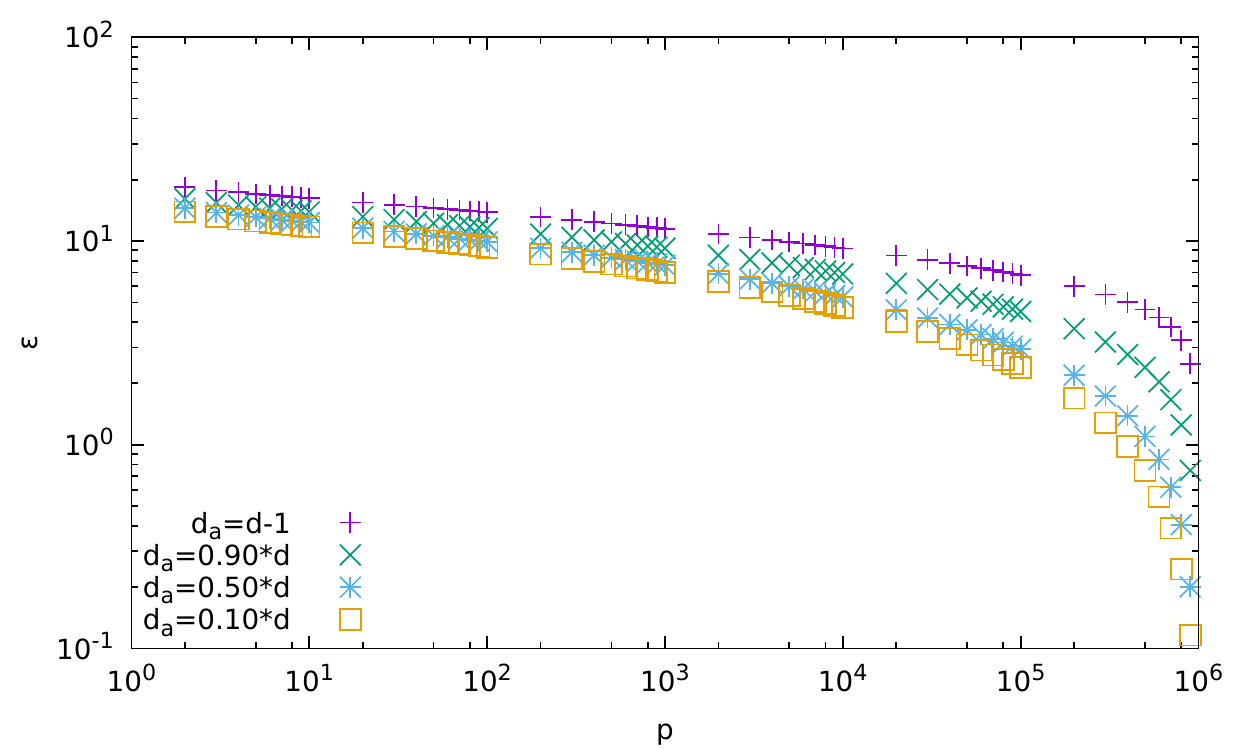}
\captionsetup{justification=centering}
\caption{Direct requests: $\epsilon$ versus $p$,\\ for $d=100$ and $n=10^6$.}
\label{fig:DR}
\end{figure}

\noindent\textbf{Practical values.} 
Fig.~\ref{fig:DR} illustrates Direct Request curves representing $\epsilon$ as a function of $p$ for different adversaries in the reference scenario of Certificate Transparency.
As millions of certificates have already been recorded in databases and hundreds of databases are supposed to be running all over the world, we have set $n=10^6$ and $d=10^2$ and assumed $p=10\cdot d$.
The security parameter $\epsilon$ starts above $10$ and slowly diminishes until nearly all of the records have been requested where the curves follow the vertical asymptote $p=n$.
If weaker adversaries decrease $\epsilon$ for any $p$, the difference becomes really noticeable only after requesting a tenth of the records.
Further, in order to achieve even a mediocre security of $\epsilon<1$, for any $d_a$, more than $\frac{9}{10}$ of the records have to be requested.
In the worst-case scenario where only one database is not colluding, we find the security parameter $\epsilon$ is approximately equal to $11.5$.
However if only half of the databases are corrupted, i.e. $d_a=\frac{1}{2}\cdot d$, we have $\epsilon \approx 7.6$.
To summarize for $n=10^6$, $d=10^2$ and $p=10\cdot d$, if $d_a =d-1$ we have $\epsilon \approx 11.5$ while if $d_a=\frac{d}{2}$, we have $\epsilon \approx 7.6$. For any $d_a$, to obtain $\epsilon < 1$, $p>\frac{9}{10}\cdot n$.

In the case of a small database system consisting of a few to tens of databases, each storing thousands of records, we set $n = 10^3$ and $d=10$.
When the adversary controls all databases but one, if the user only sends one request per database we have that $\epsilon \approx 7$ while when half of the databases are corrupted, $d_a=\frac{1}{2}\cdot d$, we have $\epsilon \approx 5.4$.
To summarize for $n=10^3$, $d=10$ and $p= d$, if $d_a =d-1$ we have $\epsilon \approx 7$ while if $d_a=\frac{d}{2}$, we have $\epsilon \approx 5.4$.\\

\vspace{3mm}
The above examples illustrate that for large databases, as the one considered in the motivating Certificate Transparency example, an adversary controlling about half the databases can extract a lot of information. Furthermore, information leakage does not diminish significantly based on the security parameter $p$, or for smaller databases. Thus we conclude the Direct Requests mechanism alone provides very weak privacy; however, we will show how its composition with an anonymity system can improve its performance.

\subsection{Anonymous direct requests}
\label{sec:asRequest}

\paragraph*{Bundled anonymous request}
\label{sec:asBundle}
We compose the \emph{direct requests} mechanism from the previous subsection with an anonymous channel. Each user, including the target user $\mathcal{U}_t$, sends a bundle of requests defined by the \emph{direct requests} PIR mechanism to databases through an anonymity system $\mathcal{AS}$.

The requests are \emph{bundled}, in that all requests from a specific user are linkable with each other, allowing this mechanism to be implemented by sending a single anonymous message through the $\mathcal{AS}$ per user. The $\mathcal{AS}$'s exit node receiving the bundle forwards the different sets of queries (as usual, encrypted by the user to each respective database) to the relevant database and anonymously returns the requested records from each database. 

The increased privacy of this scheme derives from the ability of the target user $\mathcal{U}_t$ to hide the use of the PIR system amongst $u-1$ other users. This strengthens the direct requests mechanism hiding $\mathcal{U}_t$'s query amongst $p-1$ random requests throughout $d$ servers. The adversary's task becomes harder as any bundle, out of $u$, could be the target's, and any query, out of $p$, the correct one.

\begin{algorithm}
\SetKwComment{Comment}{//}{}
\DontPrintSemicolon
\KwIn{\\
\qquad $Q$: $\quad (0 \leq Q < n)$; \\ 
\qquad $p$: $\quad (p > 1) \wedge p \equiv 0 \mod d $;}
$Req \gets \{ Q \}$;\\
\While { $|Req| < p$}
{
	$Q' \gets \textbf{random}(n)$;\\
	\If{$Q' \not \in Req$}
	{$Req \gets Req \cup {Q'}$;}
}
$Bundle \gets \{ \}$\\
\For{$1 \le i \le d$}
{
	\For{$1 \le j \le p/d$}
	{
		$Bundle_i \gets pop(Req)$
	}
	$Bundle \gets (DB_i, Bundle_i)$
}
$(index_r, rec_r) \gets \textbf{anonsendreceive}(DS, Bundle)$;\\
\KwRet $rec_Q$;
\caption{Bundled Anonymous Requests (User)}
\end{algorithm}

The database servers simply respond to bundles by returning the index and the records sought over the encrypted channel, the anonymity system forwarding the answer to the corresponding users.

\begin{secthm}
The \emph{bundled anonymous requests mechanism} is $\epsilon$-private with
$$ \epsilon = \ln \left ( \left ( \frac{d}{d-d_a} \cdot \frac{n-1}{p-1} - \frac{d_a}{d-d_a}\right )^2 + u -1 \right ) - \ln u.$$
\end{secthm}
\begin{proof}
By the application of our Composition Lemma (see below), and the security parameter of the direct requests mechanism.
\end{proof}

\noindent\textbf{Costs.}
As the only differences with the Direct request case is the Anonymity system and the bundling of the messages, we find the same values for the communication costs $C_{m}=p$ and the computation cost $C_{p}= p \cdot c_{acc}$.\\\

\paragraph*{Separated anonymous request}
\label{sec:asSeparate}

We may also compose the \emph{direct requests} mechanism (Sect.~\ref{sec:directreq}) with an anonymous channel in a different manner. Each user, including the target user $\mathcal{U}_t$, sends distinct requests defined by the \emph{direct requests} PIR mechanism to databases through an anonymity system $\mathcal{AS}$, whose queries are \emph{unlinkable} at the mix output.

The requests are \emph{separated}, in that all requests from a specific user are unlinkable with each other, allowing this mechanism to be implemented by sending separate anonymous messages through the $\mathcal{AS}$ to different databases. The $\mathcal{AS}$'s exit node receiving the message forwards it to the relevant database and anonymously returns the requested record. 

The increased privacy of this scheme derives from the ability of the target user $\mathcal{U}_t$ to hide the real query of the PIR system amongst $u \cdot (p - 1)$ other random queries. This strengthens the direct requests mechanism hiding $\mathcal{U}_t$'s query amongst $u\cdot p-1$ random requests throughout $d$ servers. 

\begin{algorithm}
\SetKwComment{Comment}{//}{}
\DontPrintSemicolon
\KwIn{\\
\qquad $Q$: $\quad (0 \leq Q < n)$; \\ 
\qquad $p$: $\quad (1 < p) \wedge p \equiv 0 \mod d $;}
$Req \gets \{ Q \}$;\\
\While { $|Req| < p$}
{
	$Q' \gets \textbf{random}(0, n)$;\\
	\If{$Q' \not \in Req$}
	{$Req \gets Req \cup {Q'}$;}
}
\ForAll{$i \in p$}
{
	$r \gets \textbf{pop}(Req)$;\\
	$(index_r, rec_r) \gets \textbf{anonsendreceive}(DS_{random(d)}, r)$;
}
\KwRet $rec_Q$;
\caption{Separated Anonymous Requests (User)}
\end{algorithm}

Since the Bundled Anonymous Requests mechanism leaks strictly more information than Separated Anonymous Request, the $\epsilon_{bundle}$ is also an upper bound of the Separated Anonymous Request.\\

\noindent\textbf{Costs.} As this method is similar with the Bundled case, we have for costs $C_{m}=p$ and $C_{p}=p \cdot c_{acc} $. However, the load on the anonymity system increases as there are $u \cdot p$ anonymous messages transmitted.\\

\begin{figure}[t]
\centering
\includegraphics[width=0.5\textwidth]{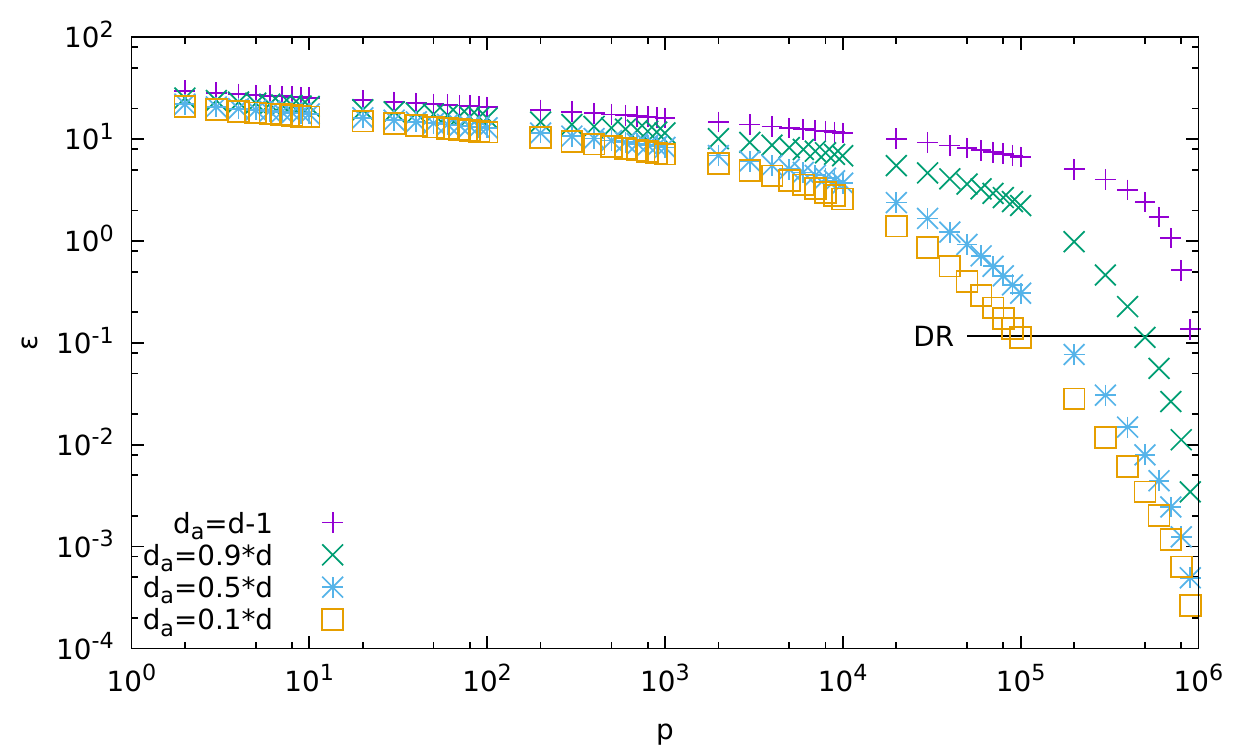}
\captionsetup{justification=centering}
\caption{AS-Bundle: $\epsilon$ versus $p$,\\ for $d=100$, $n=10^6$ and $u=10^3$.}
\label{fig:AS_DR}
\end{figure}

\noindent\textbf{Practical values.}
Fig.~\ref{fig:AS_DR} shows Direct Request composed with anonymity system curves representing $\epsilon$ as a function of $p$ for different adversaries in the reference scenario of Certificate Transparency.
As before, we set $n=10^6$ and $d=10^2$ and assumed $p=10\cdot d$ and $u=10^{3}$.
The security parameter $\epsilon$ starts above $10$ and slowly diminishes until a tenth to most of the records have been recorded depending on $d_a$ where the curves follow the vertical asymptote $p=n$.
The anonymity system gain in privacy can be seen under the line indicating where the privacy of the Direct Request protocol, without an anonymity system, stops for the same amount of points.
If the anonymity system gains appear negative at the beginning of the curves, this is due to the lack of tightness of the bound in the Composition Lemma.
If weaker adversaries decrease $\epsilon$ for any $p$, the difference becomes really noticeable only after requesting a hundredth of the records.
Further, in order to achieve even a mediocre security of $\epsilon<1$, for any $d_a$, at most half of the records have to be requested compared to 90\% without an anonymity system.
In the worst-case scenario where only one database is not colluding, we find the security parameter $\epsilon$ is approximately equal to $16$.
However if only half of the databases are corrupted, i.e. $d_a=\frac{1}{2}\cdot d$, we have $\epsilon \approx 8$.
To summarize for $n=10^6$, $d=10^2$, $u=10^3$ and $p=10\cdot d$, if $d_a =d-1$ we have $\epsilon \approx 16$ while if $d_a=\frac{d}{2}$, we have $\epsilon \approx 8$.

In the case of a small database system managing a few to tens of databases, each storing thousands of records, we again set $n = 10^3$ and $d=10$.
When the adversary controls all databases but one, each sending only one request per database, we have that $\epsilon \approx 7$ while when half of the databases are corrupted, $d_a=\frac{1}{2}\cdot d$, we have $\epsilon \approx 4$.
To summarize for $n=10^3$, $d=10$, $u=10^3$ and $p= d$, if $d_a =d-1$ we have $\epsilon \approx 7$ while if $d_a=\frac{d}{2}$, we have $\epsilon \approx 4$.\\

\vspace{3mm}
We conclude that direct requests through an anonymity system is a stronger mechanism that direct requests alone. However, for very large databases, such as the one expected in Certificate Transparency, the quality of protection is still low. It becomes better only as the total volume of requests from all users is in the order of magnitude of the number of records in the database. This requires either a large number of users, or a large number of dummy requests per user. However, even the weaker protection afforded by anonymous direct requests may be sufficient to protect privacy in applications where records only need to be accessed infrequently. 

\subsection{Sparse-PIR}
\label{sec:lPIR}

We next adapt Chor's simplest IT-PIR scheme~\cite{Chor} to reduce the number of database records accessed to answer each query. As a reminder: in Chor's scheme the user builds a set of random binary vectors of length $n$ (the number of records in the database), one for each server; we call these vectors the ``request vectors''. These are constructed so that their element-wise XOR yields a zero for all non-queried records, and a one for the record sought (we call this the ``query vector''). Each server simply XORs all records corresponding to a 1 in its request vector, and returns this value to the user. The XOR of all responses corresponds to the sought record.

Sparse-PIR aims to reduce the computational load on the database servers $\mathcal{DB}_{i}$. To this end the binary request vectors are not sampled uniformly but are sparse, requiring the database servers to access and XOR fewer records to answer each query. Specifically, in Sparse-PIR each request is derived by independently selecting each binary element using a Bernoulli distribution with parameter $\theta \leq 1/2$. Furthermore, the constraint that the XOR of these sparse vectors yields the query vector is maintained.  The intuition is that we will build a $d \times n$ query matrix $M$ \emph{column wise}: each column (of length $d$) corresponds to one record in the database, and will be selected by performing $d$ independent Bernoulli trials with parameter $\theta$, re-sampling if necessary to ensure the sum of the entries in the column (the Hamming weight) is even for non-queried records, or odd for the single queried record. 

Equivalently, we may first select a Hamming weight for each column with the appropriate probability depending on $d$, $\theta$, and whether the column represents the queried record or not, and then select a uniformly random vector of length $d$ with that Hamming weight.  Each row of the query matrix will then have expected Hamming weight $\theta \cdot n$, and the rows of the matrix (the request vectors) will XOR to the desired query vector, namely all 0 except a single 1 at the desired location.

\begin{algorithm}
\SetKwComment{Comment}{//}{}
\DontPrintSemicolon
\KwIn{\\
\qquad $Q$: $\quad 0 \leq Q < n$;\\
\qquad $\theta$: $\quad 0 < \theta \le \frac{1}{2}$;\\
}

$M \gets [\,]$;\\

\For{$0 \le col < n$}{
	\eIf{$col = Q$}
	{$q \gets$ $d$ Bernoulli($\theta$) trials with Odd sum;}
	{$q \gets$ $d$ Bernoulli($\theta$) trials with Even sum;}
	$M \gets M \textbf{ append column } q$;
}

\For{$1 \le i \le d$}
{
	$r_i \gets$ \textbf{row} $i$ \textbf{of} $M$;\\
	$resp_i \gets \textbf{sendreceive}(DB_i, r_i)$;
}
\KwRet $\bigoplus_{1 \le i \le d} resp_i$;
\caption{Sparse-PIR (User)}
\end{algorithm}

The database logic in Sparse-PIR is identical to the logic in Chor's IT-PIR: each database server receives a binary vector, XORs all records that correspond to entries with a 1, and responds with the result. In fact the database may be agnostic to the fact it is processing a sparse PIR request, aside from the reduction in the number of entires to be XORed. For $\theta < 0.5$ the costs of processing at each database is lowered due to the relative sparsity of ones, at no additional networking or other costs.

\begin{secthm} The Sparse-PIR mechanism is $\epsilon$-private with
$$\epsilon = 4 \cdot \arctanh [(1 - 2 \theta)^{(d - d_a)}],$$ 
where $\theta$ is the parameter of the Bernoulli distribution and $d - d_a$ represents the number of honest PIR servers.
\end{secthm}
\begin{proof}
See Appendix~\ref{app:proofthm2}.
\end{proof}

\noindent As expected when $\theta = 1/2$ the privacy provided by the Sparse-PIR mechanism is the same as for the perfect IT-PIR mechanism. This fact can be derived from the tight bound on $\epsilon$ by observing that $\epsilon$ equals zero when $\theta = 1/2$.
\begin{seclem} For $\theta = 1/2$, and at least one honest server, the Sparse-PIR mechanism provides perfect privacy, namely with $\epsilon = 0$.
\end{seclem}
\noindent More interestingly, as the number of honest servers increases, the privacy of the Sparse-PIR increases for any $\theta$, and in the limit becomes perfect as in standard IT-PIR:
\begin{seclem} For an increasing number of honest servers $(d - d_a) \rightarrow \infty$ the Sparse-PIR mechanism approaches perfect privacy, namely $\epsilon \rightarrow 0$.
\end{seclem}
\begin{proof}
Note that for $0 < \theta < 1$, $\lim_{x \rightarrow \infty} (1 - 2 \theta)^x = 0$. Thus for $(d - d_a) \rightarrow \infty$ we have that $\epsilon \rightarrow 0$, since $\arctanh(0) = 0$.
\end{proof}

\noindent\textbf{Costs.}
In Sparse-PIR, only $\theta \cdot n$ records are accessed and operated on per request, and $d$ of those are sent. We thus have $C_{m}=d$ and $C_{p}= \theta \cdot d \cdot n \cdot (c_{acc}+c_{prc})$.\\

\begin{figure}[t]
\centering
\includegraphics[width=0.5\textwidth]{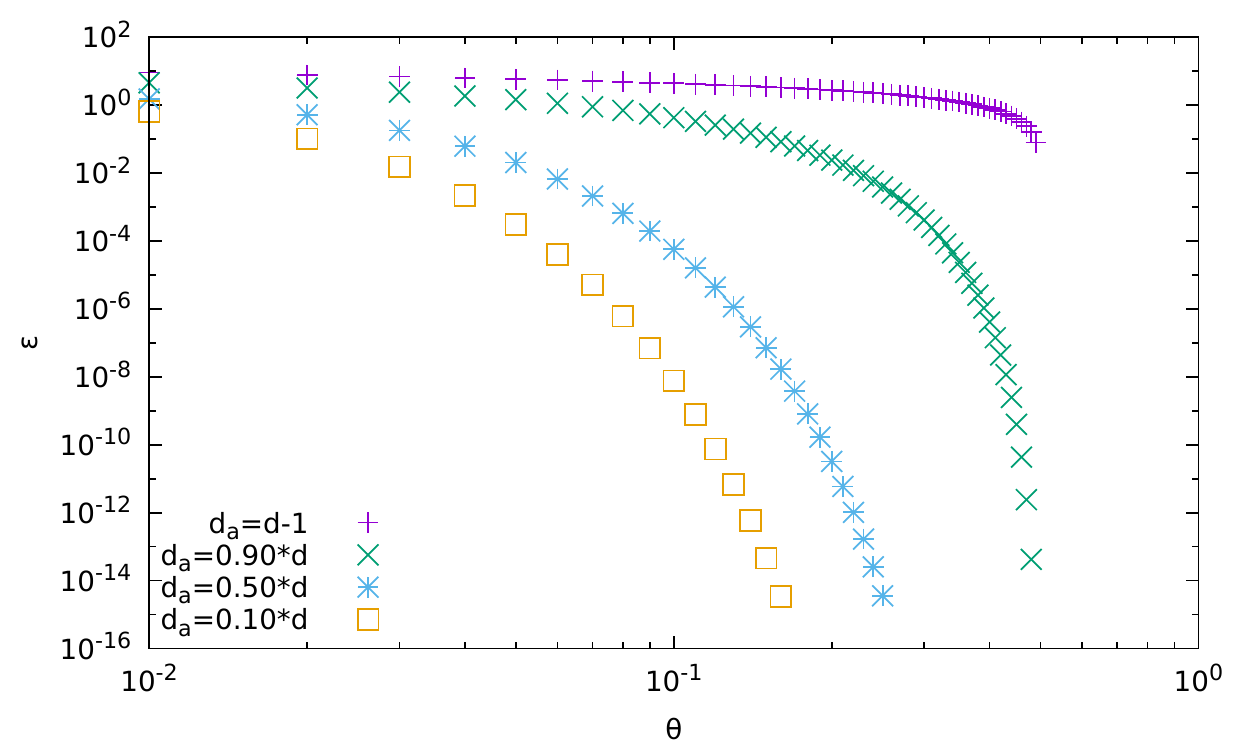}
\captionsetup{justification=centering}
\caption{Sparse-PIR: $\epsilon$ versus $\theta$ for $d=100$.}
\label{fig:SP}
\end{figure}

\noindent\textbf{Practical values.}
Fig.~\ref{fig:SP} shows Sparse-PIR curves representing $\epsilon$ as a function of $\theta$ for different adversaries in the reference scenario of Certificate Transparency with $d=10^2$.
The security parameter $\epsilon$ starts below $10$ and slowly diminishes until nearly all of the records have been accessed for $\theta=\frac{1}{2}$ where the curves follow a vertical asymptote.
The difference in $\epsilon$ for different adversaries is noticeable at any point of the curves.
In order to achieve even a mediocre security of $\epsilon<1$, except for the worst case $d_a=d-1$, accessing 10\% of the records at each database is enough.
In the worst-case scenario where only one database is not colluding, we find the security parameter $\epsilon$ is approximately equal to $2$ for $\theta = 0.25$.
However if only half of the databases are corrupted, i.e. $d_a=\frac{1}{2}\cdot d$, we have $\epsilon \approx 10^{-15}$ for the same $\theta$.
To summarize for $d=10^2$ and $\theta = 0.25$, if $d_a =d-1$ we have $\epsilon \approx 2$ while if $d_a=\frac{d}{2}$, we have $\epsilon \approx 10^{-15}$.

In the case of a small database systems managing a few to tens of databases, we set $d=10$.
When the adversary controls all databases but one, we have the $\epsilon \approx 2$ while when half of the databases are corrupted, $d_a=\frac{1}{2}\cdot d$, we have $\epsilon \approx 10^{-1}$.
To summarize for $d=10$ and $\theta = 0.25$, if $d_a =d-1$ we have $\epsilon \approx 2$ while if $d_a=\frac{d}{2}$, we have $\epsilon \approx 10^{-1}$.\\

\vspace{3mm}
A sparse version of the simple Chor scheme can indeed protect the user's privacy better than the direct request, as we can observe a factor of 9 between the two epsilons. Yet, in the worst-case scenario, where the adversary controls all the databases except one, the risk is still significant: the adversary infers that the user is about 7 times more likely to seek a particular record over another. Thus we consider strengthening the system through composition with an anonymous channel.

\subsection{Anonymous Sparse-PIR}
\label{sec:aslPIR}

We consider the composition of the Sparse-PIR mechanism with an anonymity system. In this setting, a number of users $u$ select their queries to the database servers, and perform them anonymously through an anonymity system. We consider that all requests from the same user are linkable to each other at the input and output of the anonymity system. As per our standard setting, the adversary provides a target user $\mathcal{U}_{t}$ with queries $Q_i$ and $Q_j$, one of which the user choses, and all other $u-1$ users with $Q_0$. They all use an \emph{arbitrary} $\epsilon$-private PIR mechanism through an anonymity channel to perform their respective queries.

\begin{algorithm}
\SetKwComment{Comment}{//}{}
\DontPrintSemicolon
\KwIn{\\
\qquad $Q$: $\quad 0 \leq Q < n$;\\
\qquad $\theta$: $\quad 0 < \theta \le \frac{1}{2}$;\\
}

$M \gets [\,]$;\\

\For{$0 \le col < n$}{
	\eIf{$col = Q$}
	{$q \gets$ $d$ Bernoulli($\theta$) trials with Odd sum;}
	{$q \gets$ $d$ Bernoulli($\theta$) trials with Even sum;}
	$M \gets M \textbf{ append column } q$;
}

\For{$1 \le i \le d$}
{
	$r_i \gets$ \textbf{row} $i$ \textbf{of} $M$;\\
	$resp_i \gets \textbf{anonsendreceive}(DB_i, r_i)$;
}
\KwRet $\bigoplus_{1 \le i \le d} resp_i$;
\caption{Anonymous Sparse-PIR (User)}
\end{algorithm}

We will show that this mechanism is $\epsilon$-private, through first proving a general composition lemma. This could be of independent interest to designers of private query systems based on anonymous channels.

\begin{complem}
The composition of an arbitrary $\epsilon_1$-private PIR mechanism with a perfect anonymity system used by $u$ users, for sufficiently large $u$, yields an $\epsilon_2$-private PIR mechanism with:
\begin{align*}
\epsilon_2 = \ln(e^{2\epsilon_1} + u-1) - \ln u.
\end{align*}
\end{complem}
\begin{proof}
See Appendix~\ref{app:proofcomplem}. Note this is not a worst-case analysis, but an average-case analysis. Namely there is a negligible probability in $u$, the number of users in the anonymity system, this does not hold. A fuller ($\epsilon$, $\delta$)-privacy definition could capture the worst-case behaviour.
\end{proof}

It is easy to show that as $u \rightarrow \infty$, the parameter $\epsilon_2 \rightarrow 0$, leading to a perfect IT-PIR mechanism, independently of the value of $\epsilon_1$ (so long as it is finite). Conversely, when $u = 1$, we have $\epsilon_2 = 2 \epsilon_1$ (the loss of a factor of 2 is due to the lack of tightness of the bound). Using this lemma, we can prove our main theorem.

\begin{secthm}
The composition of the Sparse-PIR scheme with parameters $\theta$, $d$, and $d_a$ with an anonymity system with $u$ users is also $\epsilon$-private with security parameter
\begin{align*}
\epsilon = \ln\left( \left( \frac{ 1 + (1 - 2 \theta)^{(d - d_a)}}
{1 - (1 - 2 \theta)^{(d - d_a)} }
\right)^4 + u - 1 \right) - \ln u.
\end{align*}
\end{secthm}
\begin{proof}
This is a direct consequence of the Composition Lemma, the security parameter of Sparse-PIR, and the definition of $\arctanh$.
\end{proof}

\noindent\textbf{Cost}
The use of an anonymity system does not change any of the server-side costs. The communication costs remain $C_{m}=d$ and the computation costs remain $C_{p}=\theta \cdot d \cdot n \cdot (c_{acc}+c_{prc})$.\\\

\begin{figure}[t]
\centering
\includegraphics[width=0.5\textwidth]{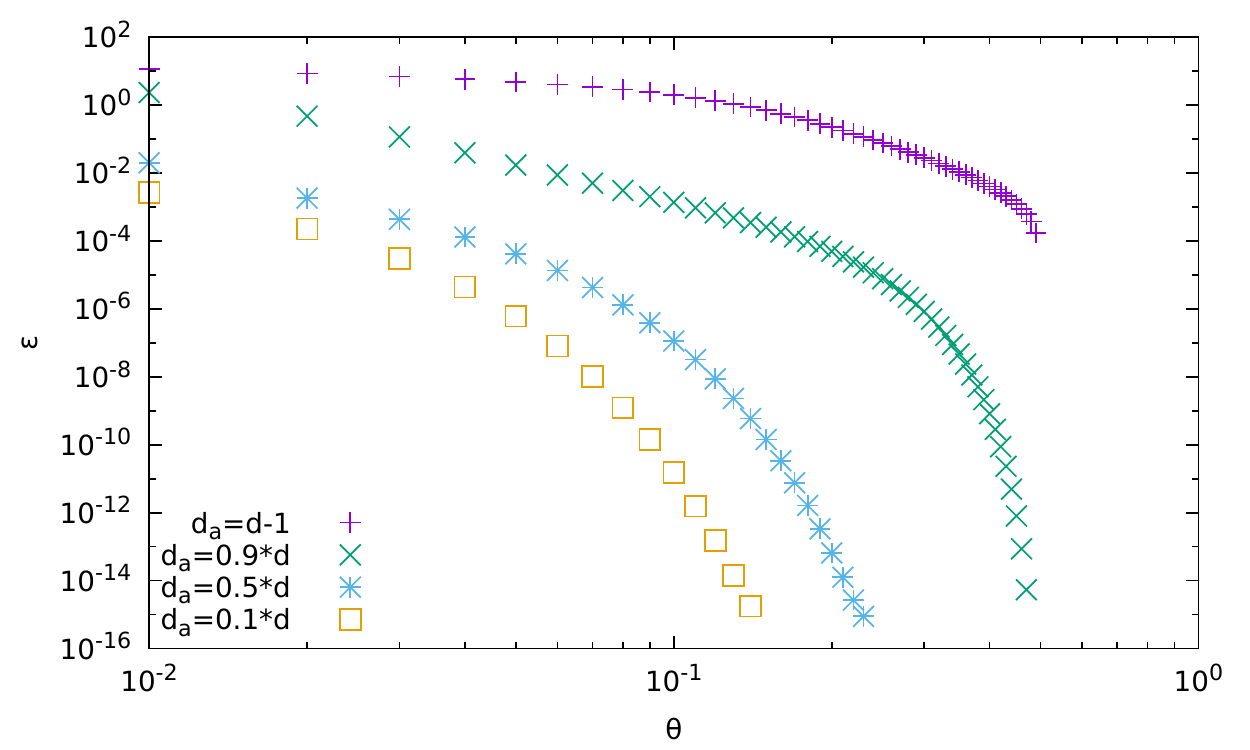}
\captionsetup{justification=centering}
\caption{AS-Sparse-PIR: $\epsilon$ versus $\theta$ for $d=100$ and $u=10^3$.}
\label{fig:AS_SP}
\end{figure}

\noindent\textbf{Practical values.}
Fig.~\ref{fig:AS_SP} shows Sparse-PIR composed with anonymity system curves representing $\epsilon$ as a function of $\theta$ for different adversaries in the reference scenario of Certificate Transparency, with $d=10^2$ and $u=10^3$.
The security parameter $\epsilon$ starts below $10$ and slowly diminishes until nearly all of the records have been accessed for $\theta=\frac{1}{2}$ where the curves follow a vertical asymptote.
If the anonymity system gains appear negative at the beginning of the curves, this is due to the lack of tightness of the bound in the Composition Lemma.
The difference in $\epsilon$ for different adversary is noticeable at any point of the curves.
In order to achieve even a mediocre security of $\epsilon<1$, except for the worst case $d_a=d-1$, accessing more than 10\% of the records at each database is enough.
In the worst-case scenario where only one database is not colluding, assuming there are $1000$ users, we find the security parameter $\epsilon$ is approximately equal to $10^{-1}$.
However, if only half of the databases are corrupted (i.e., $d_a=\frac{1}{2}\cdot d$), we have $\epsilon < 10^{-15}$.
To summarize for $d=10^2$, $u=10^3$ and $\theta=0.25$, if $d_a =d-1$ we have $\epsilon \approx 10^{-1}$ while if $d_a=\frac{d}{2}$, we have $\epsilon < 10^{-15}$.

In the case of a small database system managing a few to tens of databases we set $d=10$.
When the adversary controls all databases but one, if there are $1000$ users, each sending only one request per database, we have the $\epsilon \approx 10^{-1}$ while when half of the databases are corrupted, $d_a=\frac{1}{2}\cdot d$, we have $\epsilon \approx 10^{-3}$.
To summarize for $d=10$, $u=10^3$ and $\theta=0.25$, if $d_a =d-1$ we have $\epsilon \approx 10^{-1}$ while if $d_a=\frac{d}{2}$, we have $\epsilon \approx 10^{-3}$.\\

\vspace{3mm}
Anonymous Sparse-PIR allows us to easily trade off $\theta$ (which governs the server-side cost of the protocol) with $u$ (the number of simultaneous users of the database).  If the number of users is high, then by composing Sparse-PIR with an anonymity system, we can reduce $\theta$ and still achieve a low $\epsilon$.

\section{Optimizing PIR}
\label{sec:optimizing}

In this section, we propose an optimization for PIR systems to render them more scalable, but at a higher risk.

\subsection{Subset-PIR}
\label{sec:subPIR}

In order to lower both the communication and computation costs, when $d \gg 1$, one could consider doing IT-PIR on a subset of just $t$ of the databases. We call this optimization \textit{Subset-PIR}.

The communication and server side computation costs are thus multiplied by a factor of $\frac{t}{d}$ at the cost of a greater risk of all contacted databases being compromised. Consequently, even if an IT-PIR scheme were perfectly private, this optimization induces a non-zero probability of the adversary being able to breach it.

\begin{algorithm}
\SetKwComment{Comment}{//~}{}
\DontPrintSemicolon
\KwIn{\\
\qquad $Q$: $\quad 0 \le Q < n$;\\
\qquad $t$: $\quad 2 \leq t \leq d$;\\
}

\For{$1 \leq j \leq t-1$}
{
	{$P_j \gets$ $n$ Bernoulli($\frac{1}{2}$) trials;}
}
\Comment{$e_Q$ is the vector with all 0s except a 1 at position $Q$}
$P_t \gets \left({\oplus}_{j=1}^{t-1} P_j\right) \oplus e_Q$;\\
$DB \gets \{\}$;\\
\While{$|DB|\leq t$}
{
	server $\gets \mathbf{random}(d)$;\\
	\If{server $\notin DB$}
	{
		{$DB \gets DB \cup \{$server$\}$;}
	}
}
\For{$1 \leq j \leq t$}
{
	$resp_j \gets \textbf{sendreceive}(DB_{DB[i]}, P_j)$;
}
\KwRet $\bigoplus_{i \in t} resp_i$;
\caption{Subset-PIR (User)}
\end{algorithm}

\begin{secthm}
$Subset$-PIR is an ($\epsilon ,\delta$)-private PIR optimization with
$$ \epsilon = 0 \text{ and } \delta = \prod_{i=0}^{t-1}\frac{d_{a}-i}{d-i} $$
where $d$ is the total number of databases, of which $d_a$ are compromised and $t \leq d_a$ represents the number of PIR servers contacted. When $t > d_a$ the mechanism provides unconditional privacy.
\end{secthm}
\begin{proof} The probability of contacting $t$ databases out of which $t_{a}$ are compromised, knowing that there are in total $d_{a}$ compromised databases out of $d$ is:
$$\Pr \left ( t_{a}, t \  |\  d_{a} \right ) = \frac{{d_{a} \choose t_{a}} \cdot {d-d_{a} \choose t-t_{a}}}{{d \choose t}}$$

The probability of contacting only compromised databases is
obtained by setting $t_{a} = t$, and so is
$\frac{{d_{a} \choose t}}{{d \choose t}}$, which equals
$\prod_{i=0}^{t-1}\frac{d_{a}-i}{d-i}$ if $t \leq d_{a}$, and
$0$ if $t > d_{a}$.
\end{proof}

\noindent\textbf{Costs.}
For Subset-PIR, as we contact $t$ databases, we have $C_{m}=t$ and using a Chor-like PIR protocol we have the computation cost $C_{p} = \frac{1}{2} \cdot t \cdot n \cdot (c_{acc}+c_{prc})$.\\

\begin{figure}[t]
\centering
\includegraphics[width=0.5\textwidth]{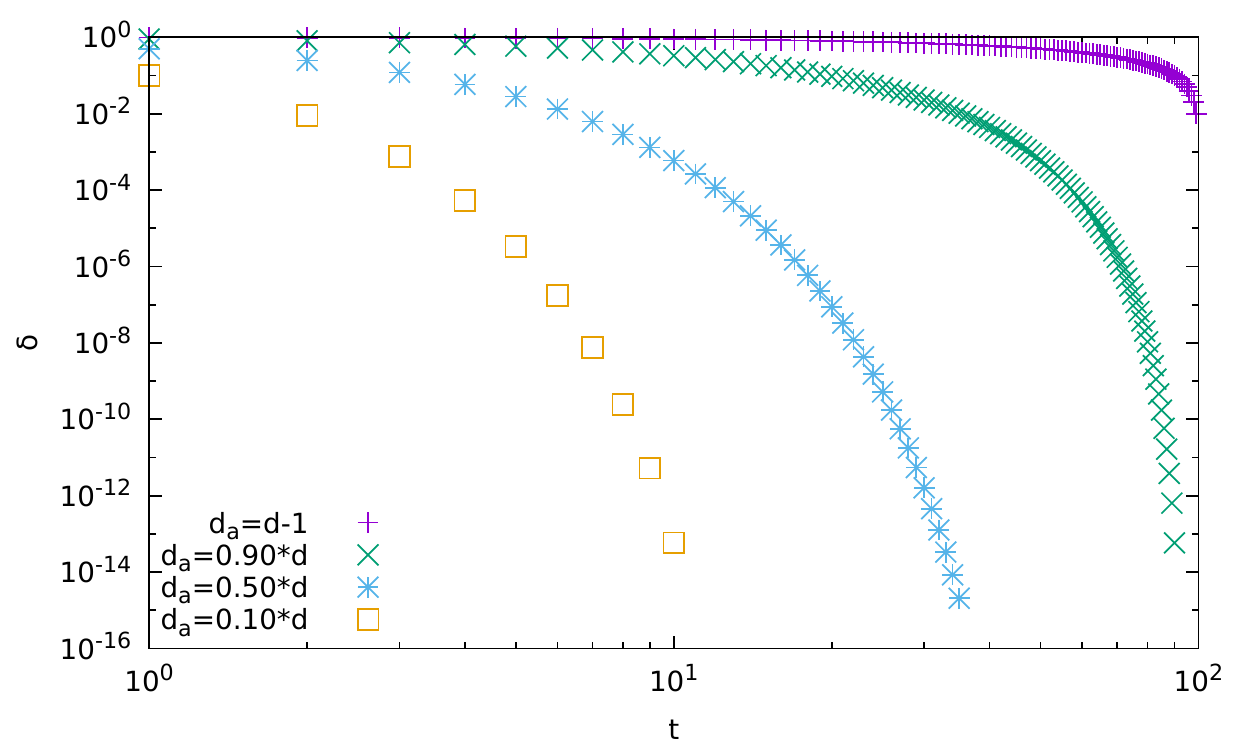}
\captionsetup{justification=centering}
\caption{Subset-PIR: $\delta$ versus $t$ for $d=100$.}
\label{fig:Sub}
\end{figure}

\noindent\textbf{Practical values.}
Fig.~\ref{fig:Sub} showns Subset-PIR curves representing $\delta$ as a function of the number of databases contacted $t$ for different adversaries in the reference scenario of Certificate Transparency, with $d=10^2$.
The security parameter $\delta$ starts between $10^{-1}$ and 1 and slowly diminishes until a tenth to most of the databases have been contacting depending on $d_a$ where the curves follow a vertical asymptote at $t=d$.
The difference in $\delta$ for different adversaries is noticeable at any point of the curves.
In order to achieve even a mediocre security of $\delta<10^{-1}$, excluding the worst case $d_a=d-1$, less than 20\% of the databases have to be contacted.
In the worst-case scenario where only one database is not colluding assuming the user contacts only a tenth of the databases, we find the security parameter $\delta$ is approximately equal to $0.9$.
However if only half of the databases are corrupted (i.e., $d_a=\frac{1}{2}\cdot d$), we have $\delta \approx 10^{-4}$.
To summarize for $d=10^2$ and $t=\frac{1}{10}\cdot d$, if $d_a =d-1$ we have $\delta \approx 0.9$ while if $d_a=\frac{d}{2}$, we have $\delta \approx 10^{-4}$.

In the case of a small database system managing a few to tens of databases, each storing thousands of records, we set $d=10$.
When the adversary controls all databases but one, if the user contacts a tenth of the databases, we have that $\delta \approx 0.9$ while when half of the databases are corrupted, $d_a=\frac{1}{2}\cdot d$, we have $\delta \approx 0.5$.
To summarize for $d=10$ and $t=\frac{1}{10}\cdot d$, if $d_a =d-1$ we have $\delta \approx 0.9$ while if $d_a=\frac{d}{2}$, we have $\delta \approx 0.5$.\\

\vspace{3mm}
Perfectly private ($\epsilon=0$) IT-PIR designs used in conjunction with the Subset-PIR optimization become ($\epsilon, \delta$)-private with $\epsilon=0$ and $\delta$ reasonably small, if the number of honest database servers is large.

\section{Comparative Evaluation}
\label{sec:comparison}

\begin{table*}[t]
\centering
\def\arraystretch{1.75}
\def\tabcolsep{0pt}
 \begin{tabular*}{\linewidth}{@{\extracolsep{\fill}}lcccc}
 \hline
 \multicolumn{1}{c}{} 			&\multicolumn{1}{c}{$\epsilon$} 															&\multicolumn{1}{c}{$\delta$} 		&\multicolumn{1}{c}{$C_m$} 	&\multicolumn{1}{c}{$C_p$} 	\\ \hline
 Chor PIR \cite{Chor}			& $0$																			&$0$					&$d$				&$\frac{1}{2} \cdot d \cdot n \cdot (c_{acc}+c_{prc})$	\\			
 Direct Requests			& $\ln \left ( \frac{1}{d-d_a}\cdot \left ( d \cdot \frac{n-1}{p-1} - d_a\right ) \right )$								&$0$					&$p$				&$p \cdot c_{acc}$ 		\\
 Sparse-PIR				& $4 \cdot \arctanh [(1 - 2 \theta)^{(d - d_a)}]$													&$0$					&$d$				&$\theta \cdot d \cdot n \cdot (c_{acc}+c_{prc})$	\\
 $\mathcal{AS}$-Request  		& $\ln \left ( \frac{1}{u} \left ( \frac{d}{d-d_a} \cdot \frac{n-1}{p-1} - \frac{d_a}{d-d_a}\right )^2 + \frac{u -1}{u} \right )$ 		&$0$					&$p$				&$p \cdot c_{acc}$ 			\\
 $\mathcal{AS}$-Sparse-PIR		& $\ln \left( \frac{1}{u} \left( \frac{ 1 + (1 - 2 \theta)^{(d - d_a)}} {1 - (1 - 2 \theta)^{(d - d_a)} } \right)^4 + \frac{u - 1}{u} \right)$		&$0$					&$d$				&$\theta \cdot d \cdot n \cdot (c_{acc}+c_{prc})$	\\
 Subset-PIR		 		& $0$																			&$\prod_{i=0}^{t}\frac{d_{a}-i}{d-i}$	&$t$				&$\frac{1}{2} \cdot t \cdot n \cdot (c_{acc}+c_{prc})$ 	\\
 \end{tabular*}
 \captionsetup{justification=centering}
 \caption{ Security and Cost Summary of the Schemes}\label{tab:summaryOpt}
\end{table*}

\begin{figure*}[t]
 \centering
 \begin{subfigure}{.5\textwidth}
  \centering
  \includegraphics[width=\textwidth]{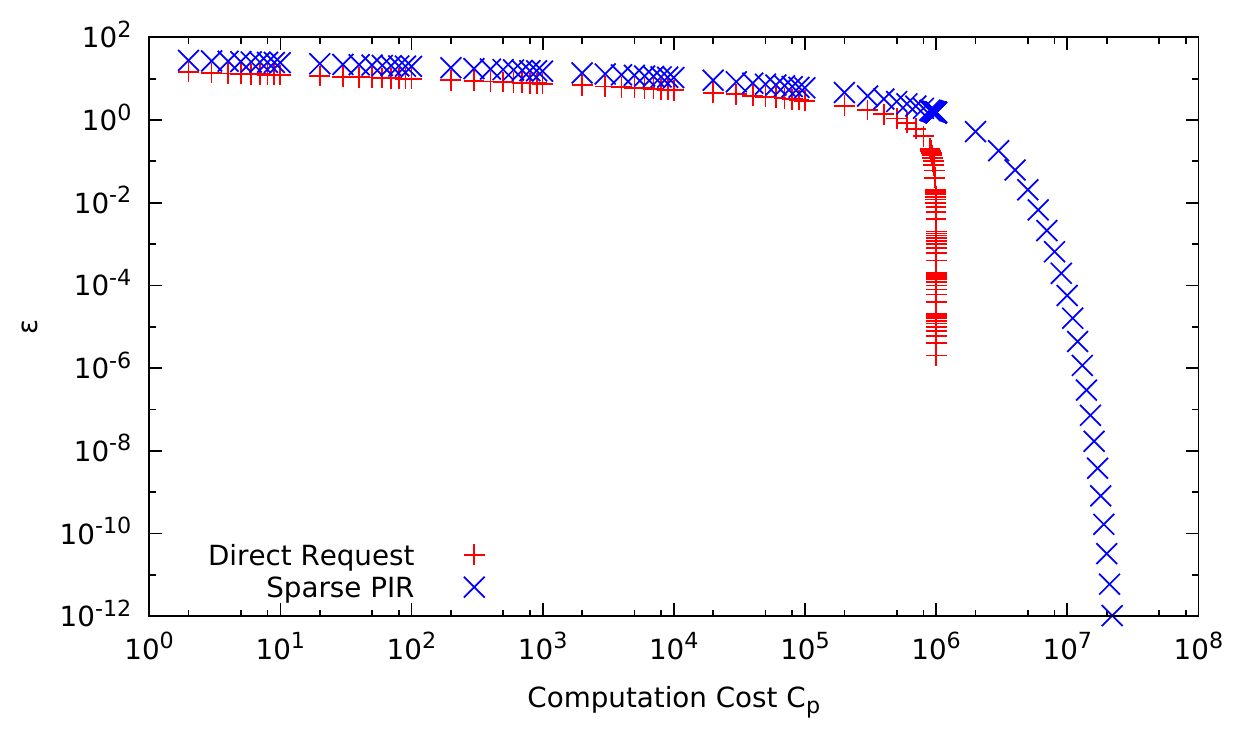}
  \captionsetup{justification=centering}
  \caption{$C_p$ versus $\epsilon$ for Direct Request ($C_p=p$)\\and Sparse-PIR ($C_p=\theta \cdot d\cdot n$).}
  \label{fig:noASCp}
 \end{subfigure}%
 \begin{subfigure}[r]{.5\textwidth}
  \centering
  \includegraphics[width=\textwidth]{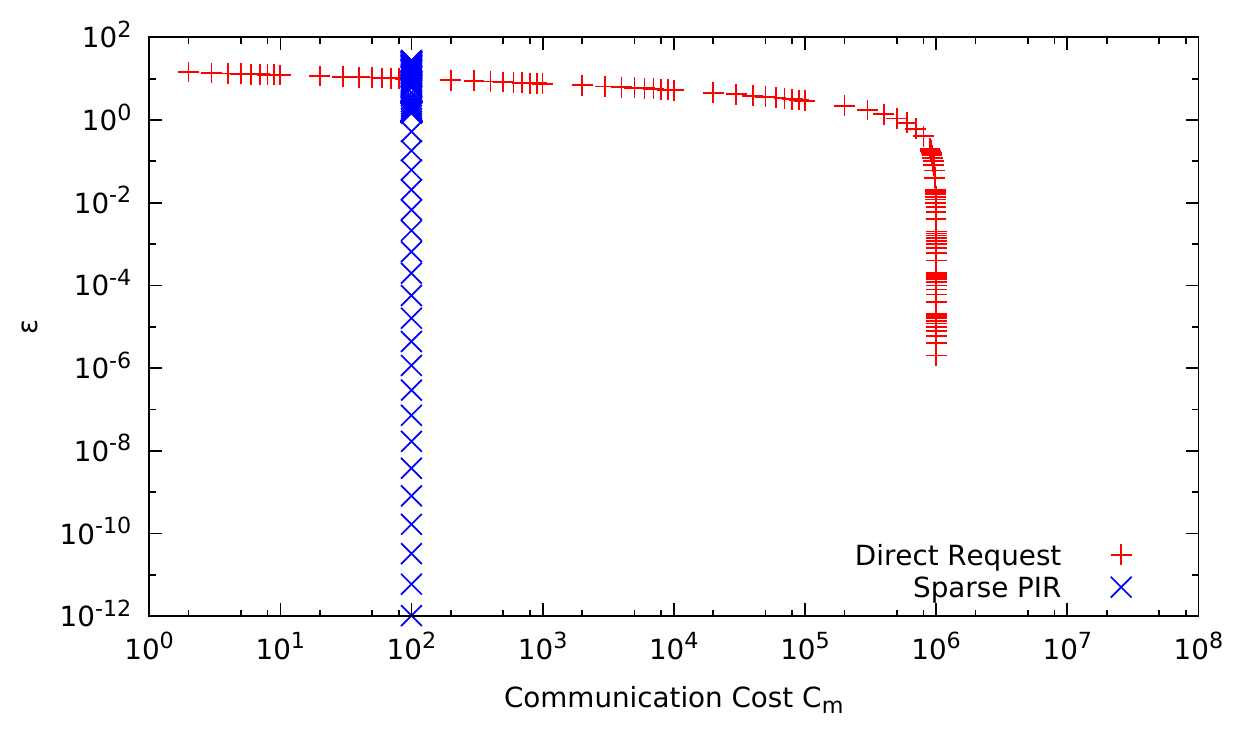}
  \captionsetup{justification=centering}
  \caption{$C_m$ versus $\epsilon$ for Direct Request ($C_m=p$)\\and Sparse-PIR ($C_m=d$).}
  \label{fig:noASCm}
 \end{subfigure}
 \begin{subfigure}{.5\textwidth}
  \centering
  \includegraphics[width=\textwidth]{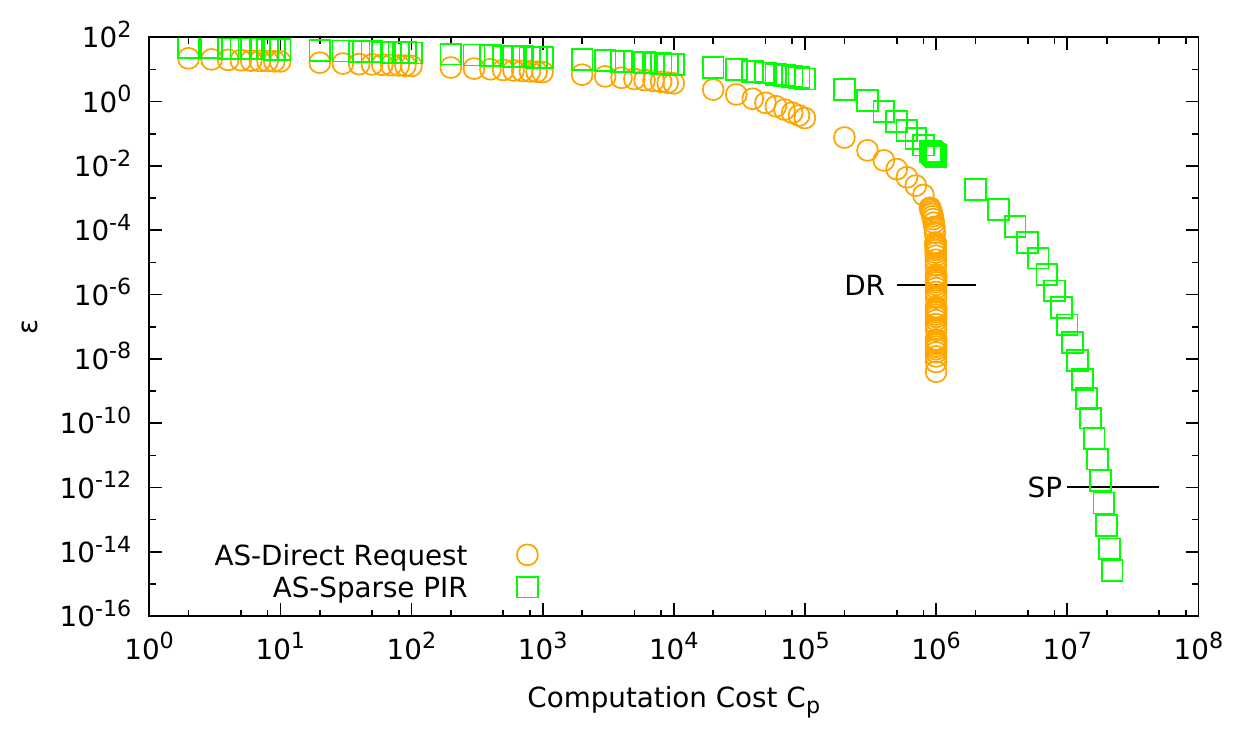}
  \captionsetup{justification=centering}
  \caption{$C_p$ versus $\epsilon$ for AS-Direct Request ($C_p=p$)\\and AS-Sparse-PIR ($C_p=\theta \cdot d\cdot n$).}
  \label{fig:ASCp}
 \end{subfigure}%
 \begin{subfigure}{.5\textwidth}
  \centering
  \includegraphics[width=\textwidth]{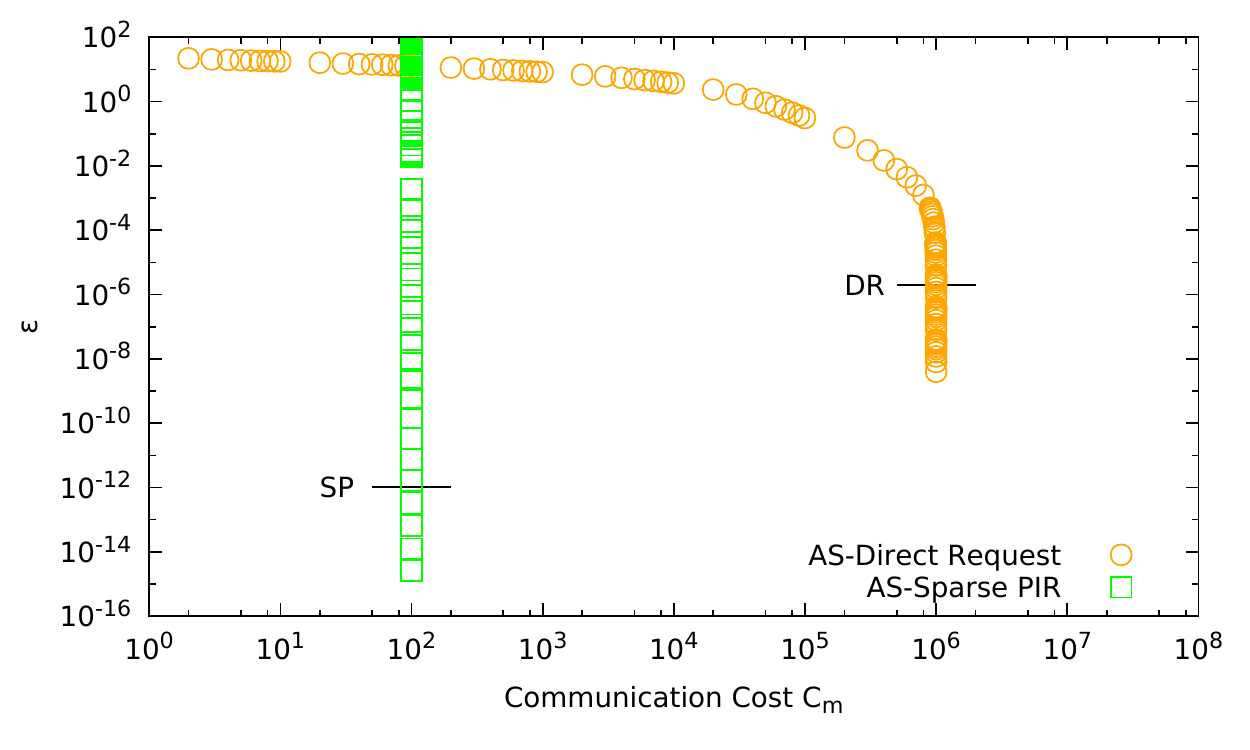}
  \captionsetup{justification=centering}
  \caption{$C_m$ versus $\epsilon$ for AS-Direct Request ($C_m=p$)\\and AS-Sparse-PIR ($C_m=d$).}
  \label{fig:ASCm}
 \end{subfigure}
 \captionsetup{justification=centering}
 \caption{Parameterized plots for Direct Request and Sparse-PIR, AS-Direct Request, and AS-Sparse-PIR, for $d=10^2$, $d_a=\frac{d}{2}$, $n=10^6$, and $u=10^3$.  The dots in the figures show the varying parameter $p$ (for the Direct Request schemes) or $\theta$ (for the Sparse-PIR schemes).}
 \label{fig:Costs}
\end{figure*}

In Table~\ref{tab:summaryOpt}, we summarize for each scheme presented the security parameters $\epsilon$ and $\delta$, the communication costs $C_{m}$, the number of blocks sent back to the user, and the computational cost $C_{p}$ which depends on the access cost $c_{acc}$ and the processing cost $c_{prc}$; i.e., the cost associated to the number of records XORed.

When the protocols are not fully private (i.e., $\epsilon \neq 0$), we observe a reduction in the server computation costs. The Sparse-PIR scheme diminishes the computation cost by a factor of $2\cdot \theta$ compared to Chor PIR~\cite{Chor}, while the Direct Request schemes induce no record processing. As the use of an anonymity system raises the privacy level, the security parameter can be lowered to reach the same privacy level of the schemes at the cost of network delays. The Sparse-PIR methods do not influence the communication cost, but the Direct Request schemes drastically increase it as the number of requests $p$ is a multiple of $d$.

The Subset-PIR optimization schemes helps scalability by reducing all costs by a factor of $\frac{t}{d}$, but turns $\epsilon$-private protocols into ($\epsilon,\delta$)-private ones.

The two main approaches for decreasing the computation are contacting fewer databases and accessing (or processing) fewer records per server. It can be noted, for example, that in order for Sparse-PIR to achieve a similar level of computation to Subset-PIR with a given $t$, the parameter $\theta$ must be particularly low, $\theta = \frac{t}{4\cdot d}$. The first approach would be relevant in the case of a quasi-trusted database system while the second in the case of a large untrusted one. 

In Figure~\ref{fig:Costs}, we compare the computation cost $C_p$, the number of records accessed, and the communication cost $C_m$, the number of records sent, of the Direct Request and Sparse-PIR schemes, and their compositions with an anonymity system, for a system comparable to Certificate Transparency when the adversary controls half of the databases.
If the costs of the designs with an anonymity system first appear greater than in the simple case, this can be explained by the lack of tightness of the bound in the Composition Lemma. The gains of the anonymity system can be seen by the values $\epsilon$ takes under the lines ``DR'' and ``SP'' which represent the last security value respectively for the Direct Request and Sparse-PIR designs without an anonymity system.

In Figures~\ref{fig:noASCp} and \ref{fig:ASCp}, we show the privacy parameter $\epsilon$ as a function of the whole database system computation cost $C_p$ and compare it between the two PIR designs and their composition with an anonymity system. For the Direct Request cases, $C_p$ represents the total number of records accessed $p$ while for Sparse-PIR ones this is the sum of the records accessed by each database $\theta \cdot d \cdot n$. This difference is worth mentioning as by definition a record can be accessed and sent only once in the Direct Request cases, while in the Sparse-PIR ones, a record can be accessed and processed at different servers. Thus, the privacy level will converge to $0$ for $p=d$ with the Direct Request protocols but for $\theta = \frac{1}{2}$, or $p=\frac{1}{2}\cdot d \cdot n$ in the graphs, with the Sparse-PIR protocols. 
While both figures show $\epsilon$ decreasing with $C_p$, the Direct Request protocols perform better for a given $C_p$ than the Sparse-PIR ones which however appear more flexible as the security parameter $\epsilon$ can be selected in a wider interval.

In Figures~\ref{fig:noASCm} and \ref{fig:ASCm}, we show $\epsilon$ as a function of the number of records sent back by the whole database system to the user and compare it between the PIR designs and their compositions with an anonymity system. While the privacy level does not depend on $C_m$ for the Sparse-PIR protocols, as the number of requests sent and record received is a constant, $C_m$ has to greatly increase to reach an adequate $\epsilon$ in the Direct Request cases.

While the Direct Requests protocols present lower computational costs than the Sparse-PIR ones, they vastly increase the communication costs. This is not a surprise as PIR was conceived in order to limit the communication cost of private search in public databases. Choosing which method to use thus depends on the database system characteristics, not only the number of database servers and the level of trust the user has, but also the hardware. One method can be used to counter the system bottleneck, Sparse-PIR would suit servers with fast processors while Direct Request would adapt better with high-speed networks. As both processing and networking capabilities are continually increasing, the question of whether Direct Request schemes have a future is still open. 

\section{Conclusions}
\label{sec:conclusions}

We show that $\epsilon$-private PIR can be instantiated by a number of systems, using dummy queries, anonymous channels, and variants of the classic Chor protocol. Yet some popular naive designs based on dummies or anonymous channels alone fail to provide even this weaker notion of privacy. We argue that the weaker protection provided by $\epsilon$-private PIR may be sufficient to provide some privacy in systems that are so large in terms of database size, but also so popular, that current IT-PIR techniques are impossible to apply.
With a large fraction of honest servers even weak (but still $\epsilon$-private) variants of PIR, such as Sparse-PIR, provide near-perfect privacy. 
Showing that a system is $\epsilon$-private enables smooth composition with an anonymity system, which guarantees that any anonymized $\epsilon$-private PIR mechanism becomes near perfect given a large enough anonymity set.

\bibliographystyle{splncs03}
\bibliography{research}

\appendix 

\section{Proofs of Theorems}

\subsection{Proof of Composing Naive Mechanisms}
\label{app:naivecompproof}
\begin{proof}
We want to prove in our indistinguishability game scenario that the probability the adversary observes exactly $u$ queries $Q_i$ is bounded above by $\delta_u \le \left ( \frac{p-1}{n-1}\right )^{u-1}$ while the probability they receive no $Q_i$ queries is bounded above by $\delta_0 \le \left ( \frac{n-p}{n-1}\right )^{u-1}$.
We first assume that the probability a user chooses one of the two queries ($Q_i$) given by the adversary is $\Pr_T$.

The probability a non-target user selects this very $Q_i$ out of his $p-1$ randomly selected requests (the $p^{th}$ one being the adversarially provided query $Q_0$) is $\frac{{n-2 \choose p-2}}{{n-1 \choose p-1}}=\frac{p-1}{n-1}$ as each record can only be requested once by any given user.
As each user is independent, the probability all the users select $Q_i$ is the product of the probabilities, we thus have $\delta_u=\Pr_T \left ( \frac{p-1}{n-1}\right )^{u-1}$.
Similarly, the probability a non-target user does not select this very $Q_i$ out of his $p-1$ randomly selected requests is $\frac{{n-2 \choose p-1}}{{n-1 \choose p-1}}=\frac{n-p}{n-1}$.
As each user is independent, the probability none of the users selects $Q_i$ is the product of the probabilities, so $\delta_0=\Pr_T \left ( \frac{n-p}{n-1}\right )^{u-1} \le \left ( \frac{n-p}{n-1}\right )^{u-1}$, and similarly for $\delta_u$.
\end{proof}

\subsection{Proof of Security Theorem 1 (Direct Requests)}
\label{app:proofthm1}
\begin{proof}
We want to prove the following result.
\begin{flalign*}
\mathcal{L} &= \frac{\mathcal{P}_{1}}{\mathcal{P}_{2}} = \frac{ \Pr (Observation \  |\   Q_{Target}=Q_{1})}{ \Pr (Observation \  |\   Q_{Target}=Q_{2})} &\\
&\leq \frac{1}{d-d_a}\cdot \left ( d \cdot \frac{n-1}{p-1} - d_a\right )  &
\end{flalign*}

We first note that the best observation for the adversary is to see exactly one of the $Q_i$, for instance $Q_1$.

In the first case, the adversary supposes $Q_1$ was sent. $Q_2$ may also have been sent, but in this case a non-colluding database would have received it.
\begin{flalign*}
\mathcal{P}_{1}&= \frac{d_a}{d} \cdot {n-1 \choose p-1}^{-1} \cdot \left [ {n-2 \choose p-1 }+ \frac{d-d_a}{d}\cdot{n-2 \choose p-2} \right ]&\\
&= \frac{d_a}{d} \cdot {n-1 \choose p-1}^{-1} \cdot \left [ {n-1 \choose p-1} - \frac{d_a}{d} \cdot {n-2 \choose p-2} \right ]&\\
&= \frac{d_a}{d} \cdot \left [ 1- \frac{d_a}{d} \cdot \frac{p-1}{n-1} \right ]&\\
\end{flalign*}

In the second case, the adversary supposes $Q_2$ was sent however she only sees $Q_1$. $Q_2$ must thus have been received by a non-colluding database.
\begin{flalign*}
\mathcal{P}_{2}&= \frac{d_a}{d} \cdot \frac{d-d_a}{d} \cdot {n-2 \choose p-2} \cdot {n-1 \choose p-1}^{-1} &\\
&= \frac{d_a}{d}\cdot \frac{d-d_a}{d} \cdot \frac{p-1}{n-1}&
\end{flalign*}

Therefore we obtain the result:
\begin{flalign*}
\mathcal{L} &= \frac{\mathcal{P}_{1}}{\mathcal{P}_{2}} \leq \frac{\frac{d_a}{d} \cdot \left [ 1- \frac{d_a}{d} \cdot \frac{p-1}{n-1} \right ]}{\frac{d_a}{d}\cdot \frac{d-d_a}{d} \cdot \frac{p-1}{n-1}} &\\
&\leq \frac{d}{d-d_a} \cdot \frac{n-1}{p-1} - \frac{d_a}{d-d_a} &\\
&\leq \frac{1}{d-d_a} \cdot \left ( d \cdot \frac{n-1}{p-1} - d_a \right )&
\end{flalign*}

This concludes the proof.
\end{proof}
\subsection{Proof of Security Theorem 2 (Sparse-PIR)}
\label{app:proofthm2}

\begin{proof}

We represent the $p$ requests sent by the user by $\{0,1\}^{1 \times n}$ vectors listed in a $d \times n$ matrix, each column representing a record and each row a request. The adversary $\mathcal{A}$ controlling only a set of the databases will only see some of the rows. $\mathcal{A}$ is interested in the number of ones in the columns, these numbers representing how many times each record has been requested.

We first note that the probability an ($d$, $\theta$)-Binomial variable is even is $\frac{1}{2} + \frac{1}{2}(1 - 2 \theta)^d$.~\cite{binom}

The adversary observes only the part of each column $v_i$ corresponding to the corrupt servers $d_a$. We call the adversary observation for column $i$, $o_i$, and the hidden part of the vector $h_i$. Without loss of generality we consider that $v_i \leftarrow o_i | h_i$ namely that the column for entry $i$ is the concatenation of the observed and the hidden part of the column. 

We denote the event the user queried for record $\alpha$ as $Q_{\alpha}$. For such a query our mechanism would set the column $\alpha$, namely $v_{\alpha}$, to have odd Hamming weight, and all other column $v_{\beta}, \beta \neq \alpha$ to have even Hamming weight.

To prove that the mechanism is differentially private we need to show that:
\begin{equation*}
\frac
{\Pr[\forall i. o_i | Q_{\alpha} ]}
{\Pr[\forall i. o_i | Q_{\beta} ]} \le e^{\epsilon}
\end{equation*}
However, each column of the query is sampled independently of all others, and thus it suffices to prove that:
\begin{equation*}
\frac{\prod_{\forall i.} \Pr[ o_i | Q_{\alpha} ]}
{\prod_{\forall i.} \Pr[o_i | Q_{\beta} ]} \le e^{\epsilon}
\end{equation*}
Since $\Pr[o_i | Q_{\alpha}] / \Pr[o_i | Q_{\beta}] = 1$ for 
$i \notin \{ \alpha, \beta \}$, this expression simplifies to:
\begin{equation*}
\frac{ \Pr[ o_{\alpha}| Q_{\alpha} ] \cdot  \Pr[ o_{\beta}  | Q_{\alpha} ]}
{ \Pr[o_{\alpha} | Q_{\beta} ] \cdot \Pr[o_{\beta} | Q_{\beta} ] } \le e^{\epsilon}
\end{equation*}
We have the following cases depending on the observed parity of $o_i$, based on the expected parity of the full, and partly unobserved, $v_i$ and $v_j$:
\begin{align*}
\Pr[ o_i \text{ odd} | Q_i ] &= \Pr[ h_i \text{ even}] \\
\Pr[ o_i \text{ even} | Q_i ] &= \Pr[ h_i \text{ odd}] = 1 - \Pr[ h_i \text{ even}] \\
\Pr[ o_j \text{ odd} | Q_i ] &= \Pr[ h_j \text{ odd}] = 1 - \Pr[ h_j \text{ even}] \\
\Pr[ o_j \text{ even} | Q_i ] &= \Pr[ h_j \text{ even}]
\end{align*}

For values of $\theta < 1/2$, it is the case that $\Pr[ h_i \text{ even}] > \Pr[ h_i \text{ odd}]$ and the differential privacy bound is minimized for:
\begin{align*}
\frac{ \Pr[ o_{\alpha} \text{ odd} | Q_{\alpha} ] \cdot  \Pr[ o_{\beta} \text{ even}  | Q_{\alpha} ]}
{ \Pr[o_{\alpha} \text{ odd} | Q_{\beta} ] \cdot \Pr[o_{\beta} \text{ even} | Q_{\beta} ] } &= \\
\frac{ \Pr[ h_{\alpha} \text{ even}] \cdot  \Pr[ h_{\beta} \text{ even} ]}
{ \Pr[h_{\alpha} \text{ odd}] \cdot \Pr[h_{\beta} \text{ odd}] } &= \\
\frac{ \Pr[ h_{\alpha} \text{ even}]^2 }
{ \Pr[h_{\alpha} \text{ odd}]^2 } &= \\
\left( \frac{ 1/2 + 1/2 (1 - 2 \theta)^{|h_i|}}
{ 1 - (1/2 + 1/2 (1 - 2 \theta)^{|h_i|}) }  \right)^2 &= \\
\left( \frac{ 1 +  (1 - 2 \theta)^{|h_i|}}
{ 1 - (1 - 2 \theta)^{|h_i|} } \right)^2 & 
\end{align*}
The value of $\epsilon$ such that this expression is bounded above by $e^{\epsilon}$ can be expressed in terms of an inverse hyperbolic tangent ($\arctanh{x} = \frac{1}{2} \ln\left ( \frac{1 + x}{1 - x} \right ) ; |x| < 1$):
\begin{align*}
\epsilon = 4 \cdot \arctanh (1 - 2 \theta)^{|h_i|} 
\end{align*}
This concludes the proof and the upper bound is tight.
\end{proof}

\subsection{Proof of the Composition Lemma}
\label{app:proofcomplem}

\begin{proof}
We consider the observations ${\mathcal{O}_0} \ldots {\mathcal{O}_{u-1}}$ as originating from the $\epsilon_1$-private PIR mechanism used by users $\mathcal{U}_0$ to $\mathcal{U}_{u-1}$ respectively. Without loss of generality we consider the target user $\mathcal{U}_t$ is $\mathcal{U}_0$. We try to determine a bound on the following quantity to prove $\epsilon$-privacy:
\begin{align*}
\frac{\Pr({\mathcal{O}_0} \ldots {\mathcal{O}_{u-1}}| Q_i, Q_0 \ldots Q_0 )}
{\Pr({\mathcal{O}_0} \ldots {\mathcal{O}_{u-1}}| Q_j, Q_0 \ldots Q_0 )} 
\leq e^{\epsilon_2}
\end{align*}
However, due to the use of the anonymity system the adversary has a uniform belief about the matching of all observations to all queries, out of the $u!$ possible matchings. Thus we have that:
\begin{align*}
\Pr({\mathcal{O}_0} \ldots {\mathcal{O}_{u-1}}&| Q_x, Q_0 \ldots Q_0 ) = \\
&= \frac{1}{u!} \sum_{i=0}^{u-1} (u-1)! \Pr(\mathcal{O}_i | Q_x) \prod_{j \neq i} \Pr(\mathcal{O}_j | Q_0) \\
&= \frac{1}{u} \sum_{i=0}^{u-1} \Pr(\mathcal{O}_i | Q_x) \prod_{j \neq i} \Pr(\mathcal{O}_j | Q_0)
\end{align*}

The quantity to be bound can therefore be re-written as:
\begin{align*}
&\frac
{\frac{1}{u} \sum_{i=0}^{u-1} \Pr(\mathcal{O}_i | Q_a) \prod_{j \neq i} \Pr(\mathcal{O}_j | Q_0)}
{\frac{1}{u} \sum_{i=0}^{u-1} \Pr(\mathcal{O}_i | Q_b) \prod_{j \neq i} \Pr(\mathcal{O}_j | Q_0)} = \\\
&\frac
{ \Pr(\mathcal{O}_0 | Q_a) \prod_{j \neq 0} \Pr(\mathcal{O}_j | Q_0) + \sum_{i \neq 0} \Pr(\mathcal{O}_i | Q_a) \prod_{j \neq i} \Pr(\mathcal{O}_j | Q_0)}
{ \sum_{i=0}^{u-1} \Pr(\mathcal{O}_i | Q_b) \prod_{j \neq i} \Pr(\mathcal{O}_j | Q_0)}
\end{align*}

We are now making a simplifying assumption: We consider that $\Pr(\mathcal{O}_i | Q_x) = \mu$ if the observation $\mathcal{O}_i$ was indeed produced by the query $Q_x$, an $\nu$ otherwise, and also $\mu > \nu$. Since the PIR mechanism is $\epsilon$-private we know that $\mu \leq e^{\epsilon_1} \nu$. This simplifying assumption holds for large numbers of $u$, since products of multiple individual $\Pr(\mathcal{O}_i | Q_x)$ will tend to be products of the average $\mu$ and $\nu$.

The quantity to be bound now reduces to:
\begin{align*}
\frac
{\mu^2 \mu^{u-2} + (u-1) \nu^2 \mu^{u-2}}
{\nu^2 \mu^{u-2} + (u-1) \nu^2 \mu^{u-2}}
 = \\
\frac
{\mu^2 + (u-1) \nu^2 }
{ u \nu^2 }
= \\
\frac {
\left( \frac{\mu}{\nu} \right)^2 + u-1  }
{ u } \leq \\ 
\frac {
\left( e^{\epsilon_1} \right)^2 + u-1  }
{ u } = \\ 
e^{\ln(e^{2\epsilon_1} + u-1) - \ln u}
\end{align*}
This concludes the proof.
\end{proof}

\end{document}